\documentclass[reprint,aps,prm,superscriptaddress]{revtex4-1}
\usepackage{graphicx}
\usepackage[version=4]{mhchem}
\usepackage[colorlinks, citecolor=blue, linkcolor=blue, urlcolor=blue, breaklinks]{hyperref}
\usepackage{siunitx, xspace, csquotes,braket}
\DeclareSIUnit[]\muB{\text{\ensuremath{\mu_{\textup{B}}}}}
\bibpunct{[}{]}{;}{n}{}{}
\newcommand{\unit}{1\!\!1}
\usepackage[svgnames]{xcolor}
\newif\ifshowcomments\showcommentstrue

\begin{document}

\title{Thermopower of the electron-doped manganese pnictide LaMnAsO}
\author{Manuel Zingl}
\affiliation{Center for Computational Quantum Physics, Flatiron Institute, 162 5th Avenue, New York, NY 10010, USA}
\affiliation{Institute of Theoretical and Computational Physics,
Graz University of Technology, NAWI Graz, 8010 Graz, Austria}
\author{Gernot J. Kraberger}
\affiliation{Institute of Theoretical and Computational Physics,
Graz University of Technology, NAWI Graz, 8010 Graz, Austria}
\author{Markus Aichhorn}
\email[]{aichhorn@tugraz.at}
\affiliation{Institute of Theoretical and Computational Physics,
Graz University of Technology, NAWI Graz, 8010 Graz, Austria}

\date{\today}

\begin{abstract}
Upon chemical substitution of oxygen with fluor, LaMnAsO has been electron-doped
in experiments, resulting in samples with remarkably high Seebeck coefficients of
around \SI{-300}{\micro\volt\per\kelvin} at room temperature and 3\% doping.
Within the framework of density functional theory plus dynamical mean-field
theory we are not only able to reproduce these
experimental observations, but can also provide a thorough
investigation of the underlying mechanisms. By considering 
electronic correlations in the half-filled Mn-$3d$ shells,
we trace the high Seebeck coefficient back to an asymmetry in the
spectral function, which is due to emergence of incoherent spectral weight
under doping and a strong renormalisation of the unoccupied states.
This is only possible in correlated systems
and cannot be explained by DFT-based band structure calculations.
\end{abstract}

\maketitle

\section{Introduction}
\label{sec:Intr}

In times of a drastic increase in energy consumption, the possibility
to convert otherwise wasted heat into electric energy through
thermoelectric devices becomes increasingly
important~\cite{snyder_complex_2008,Thermo_prog}. In addition to already
commercially-used narrow-gap semiconductors, like \ce{Pb(Se,Te)} and \ce{Bi2Te3},
research on thermoelectricity is also devoted to correlated materials, as it has
been shown in recent years that electronic correlations can have a strong influence
on the thermopower~\cite{Tomczak2018,HeldThermoelectrics,HauleThermoelectrics,arita1,arita2,Wissgott2,mravlje_thermopower_sr2ruo4}.

A class of strongly correlated materials for which experiments have
repeatedly reported high thermopowers (Seebeck coefficients) are
the manganese pnictides~\cite{Ba22,Ba21,Ba20,LaP_Yanagi,La04,La02,Sm01,Simonson2012, Ryu}.
We focus on one example out of this material class, electron-doped
\ce{LaMnAsO}, where doping can be accomplished experimentally
by a fractional replacement of the O atoms with F~\cite{La07,Ryu}.
The experimental facts to be explained by theory are the following:
The room-temperature Seebeck coefficient of polycrystalline \ce{LaMnAsO_{1-$\delta$}F_{$\delta$}}
samples changes from about \SI{-290}{\micro\volt\per\kelvin} at 3\% doping to
\SI{-190}{\micro\volt\per\kelvin} at 7\% doping, but jumps to only
\SI{-30}{\micro\volt\per\kelvin} at 10\% doping~\cite{Ryu}.

On the theoretical side fairly little is known on the origin of
the high Seebeck coefficients in manganese pnictides. Only for \ce{BaMn2As2}
the Seebeck coefficient was calculated within DFT and the constant scattering
time approximation, which resulted in values of around $\SI{-150}{\micro\volt\per\kelvin}$
for electron doping at \SI{300}{\kelvin}~\cite{Ba03}. However, the underlying
microscopic details have not been studied, yet.
Moreover, it has been pointed out in earlier works that the strong
electron-electron interaction in the nominally half-filled Mn-3d
shells is an important factor to understand the physical properties
of manganese pnictides~\cite{Ba14,LaMnPO01,Ba07,BaLa_Zingl}.
In that sense, one also needs to take into account electron-electron
interactions when studying thermoelectricity.

In this work, we use ab-initio density functional theory
electronic structure calculations coupled to dynamical mean-field theory
(DFT+DMFT)~\cite{DMFTelectronicstru} to develop a theoretical
understanding of the Seebeck coefficient in electron-doped \ce{LaMnAsO}.
Using the virtual crystal approximation (VCA) to simulate electron-doping,
not only the magnitude of the Seebeck coefficient, but also its doping
dependence can be understood from the picture of a doped correlated
insulator under the emergence of incoherent spectral weight due to
inelastic electron-electron scattering. Such a description is not possible
on the level of DFT, but requires at least local electronic correlations
in the Mn-$3d$ shells which are taken into account in DFT+DMFT. 
We emphasize that all calculations are performed without adjustable
doping-dependent parameters. The interaction parameters $U$ and $J$
are fixed to their values of our previous study~\cite{BaLa_Zingl},
where excellent agreement between theory and experiment for optical
properties has been found.

After an outline of the theoretical framework, we will
briefly review the necessary ingredients for high Seebeck coefficients
from an electronic structure point of view, before we turn to the
numerical calculations and comparisons between theoretical and
experimental results.

~~\\

\section{Methodology}
\label{sec:MethLa}

To describe the electronic structure and the transport properties of
electron-doped \ce{LaMnAsO} we carry out fully charge self-consistent
density functional theory plus dynamical mean-field theory (DFT+DMFT)
calculations using the TRIQS/DFTTools
package~\cite{TRIQS, TRIQS/DFTTools, LaFe01, TRIQS/DFTTools2}
in combination with WIEN2k~\cite{Wien2k1,Wien2k}. In addition to the
following outline we refer the reader to our previous work~\cite{BaLa_Zingl}
for further computational details.

For the DFT calculation we use \num{10000} $k$-points in the full Brillouin 
zone (BZ) and the standard Perdew-Burke-Ernzerhof (PBE)~\cite{PBE2} generalized 
gradient approximation (GGA) for the exchange-correlation functional.
We use the crystal structure of the undoped compound (measured at \SI{290}{\kelvin}
in Ref.~\cite{La08}) for all calculations. A test calculation with the experimental
crystal structure at 10\% electron doping~\cite{Ryu} showed no substantial changes
in our results. For all magnetic calculations we consider the C-type antiferromagnetic
(AFM) ordering as determined experimentally for the undoped compound~\cite{La06}.
We treat electron doping using the virtual crystal approximation (VCA) in DFT by modifying the atomic numbers of the substituted atoms according to the desired doping levels. We assess the quality of
this approximation by a comparison to super-cell calculations (see
appendix~\ref{sec:vca}). The doping is of course taken into account
also in the DMFT part of the calculation by adjusting the chemical
potential to the corresponding electron count.

From the DFT Bloch states we construct projective Wannier functions for the Mn-$3d$ 
orbitals in an energy window from \SI{-5.50}{\electronvolt} to \SI{3.25}{\electronvolt}
around the Fermi energy for the undoped compound, but adjust the upper boundary such
that the same number of states at each doping level are included.
In DMFT we work with a full rotationally invariant Slater Hamiltonian for the five 
Mn-$3d$ orbitals with a Coulomb interaction $U=F^0$ of \SI{5.0}{\electronvolt} 
and a Hund's coupling $J=(F^2+F^4)/14$ of \SI{0.9}{\electronvolt}
($F^4/F^2 = 0.625$)~\cite{BaLa_Zingl}. We choose the fully localized limit (FLL) as
double counting correction~\cite{FLL}, where we use the electron
charge in the $3d$ orbitals calculated from the fully
self-consistently determined charge density.
The TRIQS/CTHYB solver~\cite{TRIQS/CTHYB}, which
is based on continuous-time quantum Monte Carlo in the hybridization expansion~\cite{CTQMC,werner2},
is used to solve the impurity model on the Matsubara axis at an inverse temperature
$\beta= \SI{40}{\electronvolt^{-1}}$, corresponding to room temperature. We use the Beach's
stochastic method~\cite{MaxEntBeach} for the analytic continuation of the self-energy to the
real-frequency axis by constructing an auxiliary Green's
function $G_{aux}(z) = (z-\Sigma(z)+\Sigma(\infty)+\mu)^{-1}$.

Transport properties are evaluated within the linear response Kubo formalism
(neglecting vertex corrections). The static conductivity tensor $\sigma_R^{\alpha\alpha'}$
and the Seebeck tensor $S^{\alpha\alpha'}$ are given by~\cite{Mahan,oudovenko}
\begin{equation}
\label{eq:con_seebeck}
\sigma^{\alpha\alpha'} = K_0^{\alpha\alpha'}  \  \ \
\text{and} \ \ \  S^{\alpha\alpha'} =
- \left(K^{-1}_0\right)^{\alpha\gamma} K_1^{\gamma\alpha'},
\end{equation}
with $\alpha,\alpha',\gamma \in \{x,y,z\}$ and kinetic coefficients
\begin{equation}
\label{eq:kinetic_coeff}
   K_n^{\alpha\alpha'} = N_{\sigma} \pi \int{d\omega\
     \left(\beta\omega\right)^n
     \left(-\frac{\partial f\left(\omega\right)}{\partial \omega}\right)\Gamma^{\alpha\alpha'}\left(\omega,\omega\right)}\ , 
\end{equation}
where $N_\sigma$ is the spin degeneracy and $f(\omega)$ the Fermi function.
The transport distribution is defined as 
\begin{equation}
\label{eq:transportdistribution}
   \Gamma^{\alpha\alpha'}\left(\omega\right) = \frac{1}{V} \sum_k
   \mathrm{Tr}\left[v^\alpha(\mathbf{k})A(\mathbf{k},\omega)v^{\alpha'}(\mathbf{k})A(\mathbf{k},\omega)\right]
\end{equation}
with the unit cell volume $V$. In multi-band systems the interacting (correlated)
spectral function $A_{\nu\nu'}\left(\mathbf{k},\omega\right)$ and the velocities
$v_{\nu\nu'}^\alpha(\mathbf{k})$ are Hermitian matrices in the band indices
$\nu,\nu'$, which we omitted in the equations above. The velocities
(matrix elements of the momentum operator) are calculated with the WIEN2k \emph{optic}
code~\cite{AmbroshDraxl2006},
$v^\alpha_{\nu\nu'}(\mathbf{k}) = -i \bra{\psi_\nu(\mathbf{k})} \nabla^\alpha \ket{\psi_{\nu'}(\mathbf{k})}/m_e $,
from the charge self-consistent Bloch states.

For a crystal symmetry demanding diagonal rank-2 tensors, as it is
the case for \ce{LaMnAsO}, the Seebeck coefficient in direction
$\alpha$ is given by  
\begin{equation}
S^{\alpha} = - \frac{K_1^{\alpha\alpha}}{K_0^{\alpha\alpha}} = -\frac{K_1^{\alpha\alpha}}{ \sigma^{\alpha\alpha}}\ .
\end{equation}

As all synthesized samples of \ce{LaMnAsO} (doped and undoped) are
polycrystalline, we simulate a \enquote{polycrystalline} Seebeck
coefficient by averaging over the three Cartesian
coordinates~\cite{Sav} 
\begin{equation}
\label{eq:sav}
S_{av} = \frac{S^{xx}\sigma^{xx} + S^{yy}\sigma^{yy} + S^{zz}\sigma^{zz}}{\sigma^{xx}+\sigma^{yy}+\sigma^{zz}}\ .
\end{equation}

\begin{figure}[t]
	\centering
	\includegraphics[width=1\linewidth]{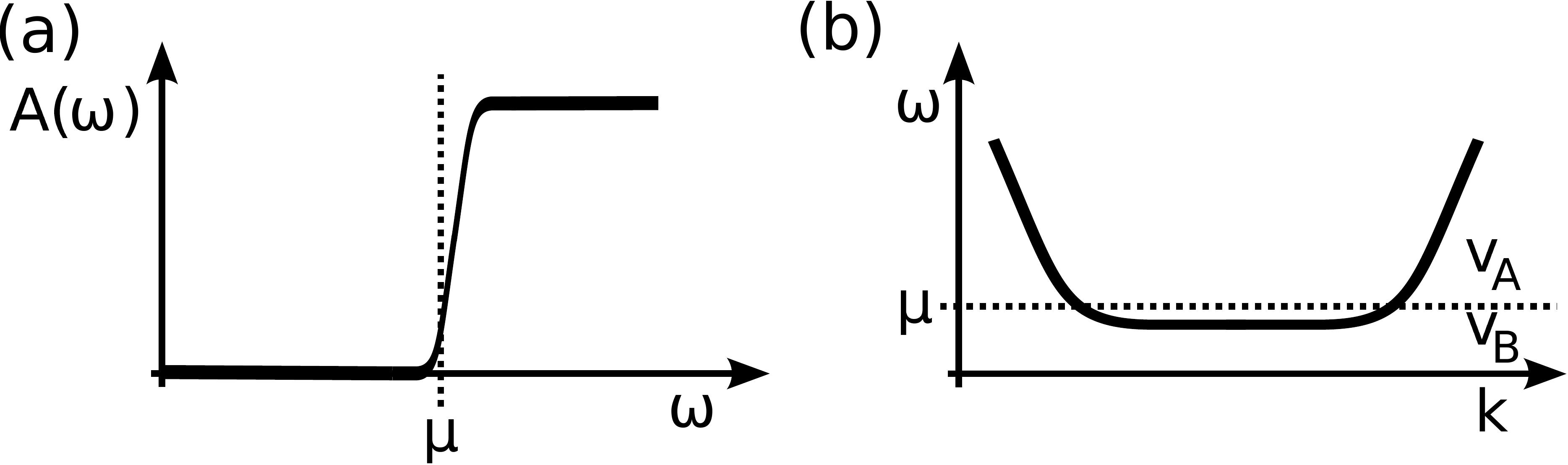}
	\caption{\label{fig:scenarios}Sketch of two scenarios promoting a high Seebeck coefficient.
		(a): \mbox{Particle-hole} asymmetry in the spectral function $A(\omega)$. (b): Asymmetry
		in the velocities $v_A \gg v_B$ due to a high dispersion above the chemical potential $\mu$
		and the flat portion of the band below, resp.}
\end{figure}

~~\\
~~\\

\section{Ingredients for a high \\Seebeck coefficient}
\label{sec:Micropic}

Due to the $\omega$-factor in the kinetic coefficient
$K_1^{\alpha\alpha'}$, Eq.~(\ref{eq:kinetic_coeff}), electron
contributions ($\omega > 0$ ) and hole contributions 
($\omega < 0$) influence the Seebeck coefficient $S$ in an opposite
way. Thus, getting a high $K_1^{\alpha\alpha'}$, and in turn a
high $S$, requires a high electron-hole asymmetry around the chemical
potential $\mu$, i.e., a strongly asymmetric transport distribution
$\Gamma^{\alpha\alpha'}$, Eq.~(\ref{eq:transportdistribution}).
There are two different mechanisms to promote a strong asymmetry in
$\Gamma^{\alpha\alpha'}$, as there are likewise two quantities
entering $\Gamma^{\alpha\alpha'}$: the velocity matrices
$v^\alpha(\mathbf{k})$ and the spectral function $A(\mathbf{k},\omega)$.  

The first scenario, shown in sketch Fig.~\ref{fig:scenarios}~(a), is to
have an asymmetric spectral function, e.g.\ with a steep slope of the spectrum
close to the chemical potential $\mu$~\cite{S_DOS1,S_DOS2,TomczakNarrowGap}.
A high positive Seebeck coefficient is expected if there is much more
spectral weight below $\mu$ than above, and a high negative Seebeck coefficient
for cases with much more spectral weight above $\mu$ than below.
In the context of strongly correlated systems this picture was also discussed
with regard to a sharp Kondo peak directly above or below the chemical
potential~\cite{HeldThermoelectrics, HauleThermoelectrics}. 

In the second scenario a high Seebeck coefficient is obtained from a
strong asymmetry directly in $v(\mathbf{k})$, which can occur due to
peculiar band shapes~\cite{arita1,arita2,arita3}.  If we assume a constant
isotropic scattering time $\tau_s$, an approximation for the kinetic
coefficients is~\cite{arita1} 
\begin{align}
  K_0 &\sim \sum_{\mathbf{k}} \left(v^2_{A}(\mathbf{k}) +
        v^2_{B}(\mathbf{k})\right), \\
  K_1 &\sim \sum_{\mathbf{k}} \left(v^2_{A}(\mathbf{k}) - v^2_{B}(\mathbf{k})\right).
\end{align}
Here, the summation runs only over states in the range of $|\omega-\mu| \lesssim 1/\beta$. 
The velocities $v_{A}$ are characteristic velocities for the states
above (A) $\mu$ and $v_{B}$ for states below (B) $\mu$. For example, a
linear dispersion in the vicinity of
$\mu$ corresponds to $v^2_{A} \approx v^2_{B}$, and thus $K_1$ will be
small, as it is the case for ordinary metals~\cite{arita1}.  
The optimal situation for a high $S$ are \enquote{pudding-mold}-like
bands, which are for instance non or only weakly dispersive below
$\mu$ and show a strongly-dispersive behavior above $\mu$, see sketch
Fig.~\ref{fig:scenarios}~(b). If $\mu$ is located close to the
flat portion of such a band and the temperature is high enough, we find
$v^2_{A} \gg v^2_{B}$, and consequently $K_1 \sim v^2_{A}$. For a band
with its flat portion below $\mu$ this results in a negative $S$.

Of course, $v(\mathbf{k})$ and $A(\mathbf{k},\omega)$ are intertwined,
and for real materials the influence of the electronic structure on
the thermoelectric properties should be always considered as an
interplay of these two ingredients~\cite{Wissgott1, Wissgott2}.
A band structure showing a strong asymmetry in $v(\mathbf{k})$ usually comes
with an asymmetry in $A(\mathbf{k},\omega)$, too.
This can be such that it partially compensate the effect of the asymmetry in
$v(\mathbf{k})$, as demonstrated for \ce{Na_{0.7}CoO2}~\cite{Wissgott1}.
We show below that also the opposite behavior, where $A(\mathbf{k},\omega)$
and $v(\mathbf{k})$ contribute with the same sign to $S$, is possible.
Coming back to the sketch in Fig.~\ref{fig:scenarios}~(b), this happens when the 
corresponding asymmetry in $A(\mathbf{k},\omega)$ is such that the derivative of
the Fermi function in the kinetic coefficient (Eq.~(\ref{eq:kinetic_coeff})) 
picks up more spectral weight above $\mu$ than below. Then, the asymmetries in 
$A(\mathbf{k},\omega)$ and $v(\mathbf{k})$ contribute both with a negative sign to $S$.

An indicator, which we will use in this work, to measure the
influence of the asymmetry in the spectral function on $S$, is to
evaluate Eqs.~(\ref{eq:con_seebeck}-\ref{eq:transportdistribution}) with
$v^\alpha(\mathbf{k}) = \text{const.} \times \unit$. In this case, the
velocities drop out and we end up with 
\begin{equation}
\label{eq:sv1}
S^{v=\unit} = -\frac{\int{d\omega\
\beta\omega\left(\partial f\left(\omega\right) / \partial \omega\right) \sum_{\mathbf{k}}
\mathrm{Tr}\left[A^2(\mathbf{k},\omega)\right]}}
{\int{d\omega\ \left(\partial f\left(\omega\right) / \partial \omega\right) \sum_{\mathbf{k}}
\mathrm{Tr}\left[A^2(\mathbf{k},\omega)\right]}}\ .
\end{equation}
If $S^{v=\unit}$ is significantly different from $S$, we can infer
that the asymmetry in the velocities is important for the Seebeck
coefficient. If $S^{v=\unit}\approx S$, the velocities are of less
importance, and a high Seebeck coefficient is driven mainly by the asymmetry
in the spectral function.

~~\\
~~\\

\section{Results}
\label{sec:ResuLa}

\begin{figure}[t]
	\centering
	\includegraphics[width=1.0\linewidth]{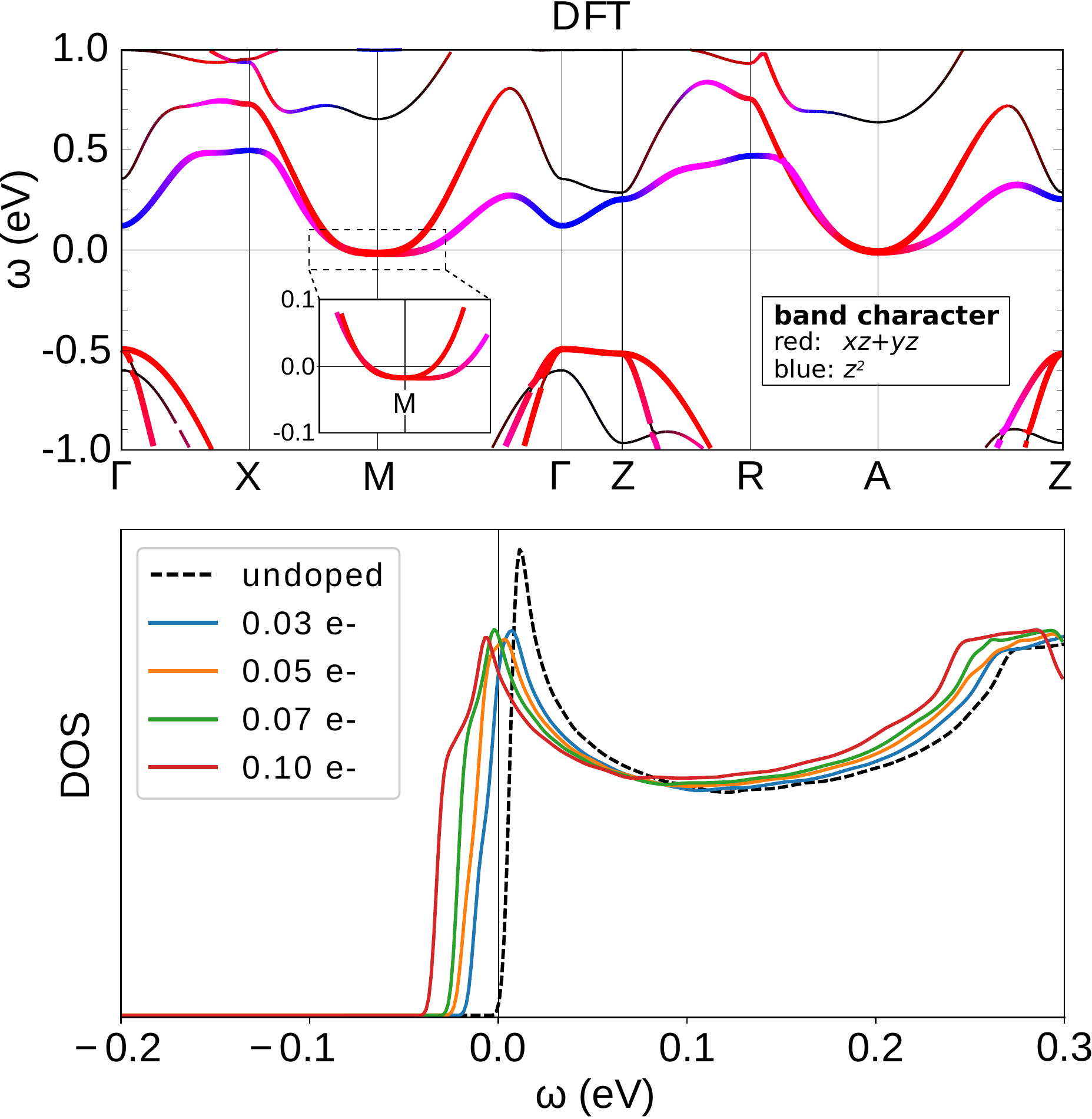}
	\caption{\label{fig:DFT}Top: Spin-polarized (antiferromagnetic) DFT band structure at
		5 \% electron doping on the $\Gamma$-X-M-$\Gamma$-Z-R-A-Z $\mathbf{k}$-path. The
		band character of the $xz/yz$ orbitals is colored in red and the $z^2$ orbitals
		are colored in blue, resp. The inset shows the low-energy region around the M point. The prima WIEN2k add-on~\cite{primapy} was used to create
		this panel. Bottom: Evolution of total antiferromagnetic DFT DOS for 3, 5, 7 and 10\%
		electron doping (colored lines). Additionally, the undoped DFT DOS (dashed black line)
		is shown, shifted such that the onset of the unoccupied states is at the Fermi
		energy ($\omega = \SI{0}{eV}$).\\}
\end{figure}

We start with the discussion of the Seebeck coefficient $S_{av}$ of \ce{LaMnAsO}
at 5\% electron doping, which was experimentally determined to be
$\SI{-240}{\micro\volt\per\kelvin}$~\cite{Ryu}. First, we calculate $S_{av}$
directly from spin-polarized (antiferromagnetic) DFT assuming a constant
isotropic scattering time $\tau_s$ (const.-$\tau_s$ approximation)~\footnote{In
the TRIQS/DFTTools transport code we achieve this by setting
$\Sigma(\omega) = -i / \tau_s$ with  $1/\tau_s = \SI{0.05}{\electronvolt}$.
Note that the actual value of $\tau_s$ is not relevant as it cancels
in the calculation of $S$.}. 
This approximation results in a high negative
value of $S_{av} = \SI{-170}{\micro\volt\per\kelvin}$. On the contrary, when
setting the velocity matrices $v(\mathbf{k}) = \unit$ (Eq.~(\ref{eq:sv1}))
we only obtain $S^{v=\unit}_{av} = \SI{-70}{\micro\volt\per\kelvin}$, showing
the importance of the asymmetry in $v(\mathbf{k})$. This is also
apparent in the DFT band structure (Fig.~\ref{fig:DFT}, top). 
The DFT picture is that of a band insulator which
becomes metallic under doping as the Fermi energy is \enquote{shifted} into the
unoccupied states. The doping mainly affects the hole pockets of
$xz/yz$ orbital character at the A and M points. At the M point
the bands are rather flat, resembling a mold-like shape with the
bottom of the $xz/yz$ bands lying below the Fermi energy. This is the origin of
the strong influence of $v(\mathbf{k})$ on $S_{av}$. The
associated asymmetry in the density of states (DOS), see bottom panel
of Fig.~\ref{fig:DFT}, does not compensate the $v(\mathbf{k})$ asymmetry,
but rather gives a contribution with the same sign and implies that at
room temperature more relevant states above the Fermi energy than
below contribute to $S_{av}$. We conclude that on the DFT level the
asymmetries in the DOS and $v(\mathbf{k})$ are both of similar relevance for $S_{av}$.

This picture drastically changes when we calculate $S_{av}$ from
the antiferromagnetic (AFM) DFT+DMFT solution instead of using
the DFT+const.-$\tau_s$ approximation. The resulting $S_{av}$
is $\SI{-230}{\micro\volt\per\kelvin}$, which is in remarkably good
agreement with the experimental value. Setting $v(\mathbf{k}) =
\unit$ leads to only a slight reduction of $|S_{av}|$ by about
15\%. In sharp contrast to the DFT+const.-$\tau_s$ result, this
reveals that the asymmetry in the DFT+DMFT spectral function is
the major factor, whereas the influence of the velocity asymmetries is negligible.

\begin{figure}[t]
	\centering
	\includegraphics[width=1.0\linewidth]{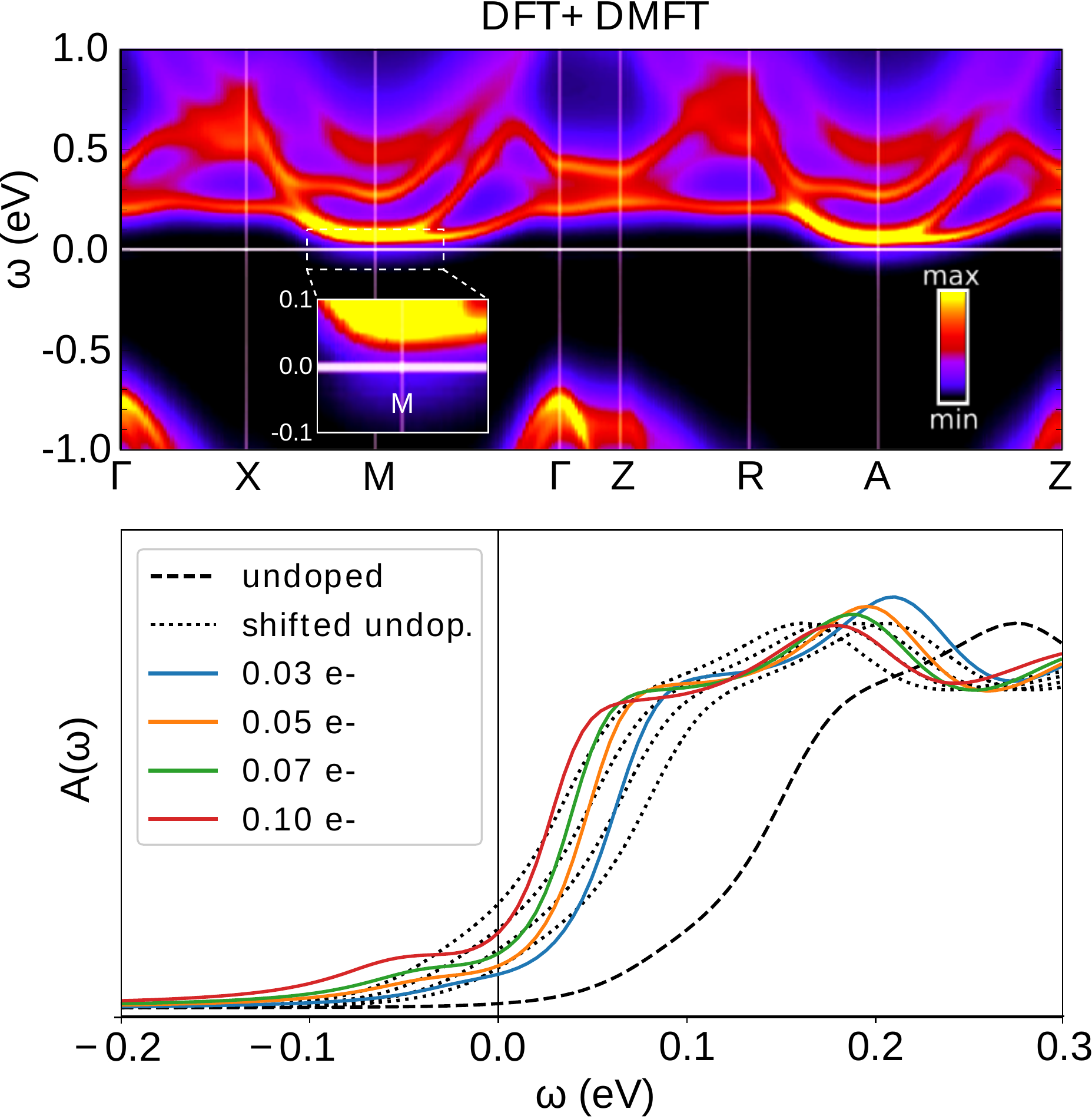}
	\caption{\label{fig:DMFT}Top: Antiferromagnetic correlated spectral function
		$A(\mathbf{k},\omega)$ at 5 \% electron doping on the $\Gamma$-X-M-$\Gamma$-Z-R-A-Z $\mathbf{k}$-path. The inset shows the low-energy region around the M point.
		Bottom: Evolution of momentum-integrated DFT+DMFT spectral functions for 3, 5, 7 and 10\% electron
		doping (colored lines). The undoped spectral function (dashed black line) is shown for comparison
		and a simple energy shift of it, adjusted to fillings corresponding to the four doping levels,
		is indicated by the dotted black lines.\\~~\\~~\\}
\end{figure}

To gain a better understanding of this observation, we
discuss how electronic correlations shape the spectral function
under doping (Fig.~\ref{fig:DMFT}, bottom). In comparison to undoped
\ce{LaMnAsO} (dashed line), doping has two major effects:
First, the edge of the spectral function between \SI{0.0}{\electronvolt} and
\SI{0.25}{\electronvolt} in the undoped compound is pushed towards
$\omega=\SI{0.0}{\electronvolt}$ with increased doping. Although the
slope of this edge is steeper in comparison to the undoped spectral
function, it does not substantially change for the different doping
levels. Second, the insulating state is suppressed and spectral weight at
and below the chemical potential emerges. The spectral weight below the
chemical potential increases with doping level and develops into a shoulder,
well visible at 10\% doping. We emphasize that this spectral weight
is not a result of the displacement of quasi-particle states, but is entirely
incoherent and originates from inelastic scattering due to electronic correlations.

~~\\
This is also visible in the $\mathbf{k}$-resolved spectral function
(Fig.~\ref{fig:DMFT}, top), where the chemical potential cuts trough incoherent
spectral weight of the unoccupied states. However, the most prominent spectral
features at the A and M points lie above the chemical potential, which leads
to a considerable spectral weight asymmetry. This is clearly different from
the DFT band structure,  where the coherent bands cross the Fermi energy
(see Fig.~\ref{fig:DFT}, top). Additionally, around the A and M points we
find spectral weight which has a stronger pronounced flat part than the
corresponding DFT bands, cf. the path from M to $\Gamma$ in the top panels
of Figs.~\ref{fig:DFT} and~\ref{fig:DMFT}. In general, spectral features are
much less dispersive in the DFT+DMFT result as a consequence of the overall
renormalization of the unoccupied states.

Additionally, we show in the bottom panel of Fig.~\ref{fig:DMFT} the
spectral functions which are generated by a rigid shift of the undoped spectral
function according to the different doping levels (dotted black lines).
The fact that not much spectral weight is present in the first
\SI{0.05}{\electronvolt} above the chemical potential results in a
substantial shift already at the lowest doping level of 3\%. A further
increase of the doping leads to only small additional shifts.
However, these simple energy shifts do not correctly reproduce the doped
DFT+DMFT results, demonstrating the importance of separate fully
charge self-consistent DFT+DMFT calculations at each doping level.

\begin{figure}[t]
	\centering
	\includegraphics[width=1.0\linewidth]{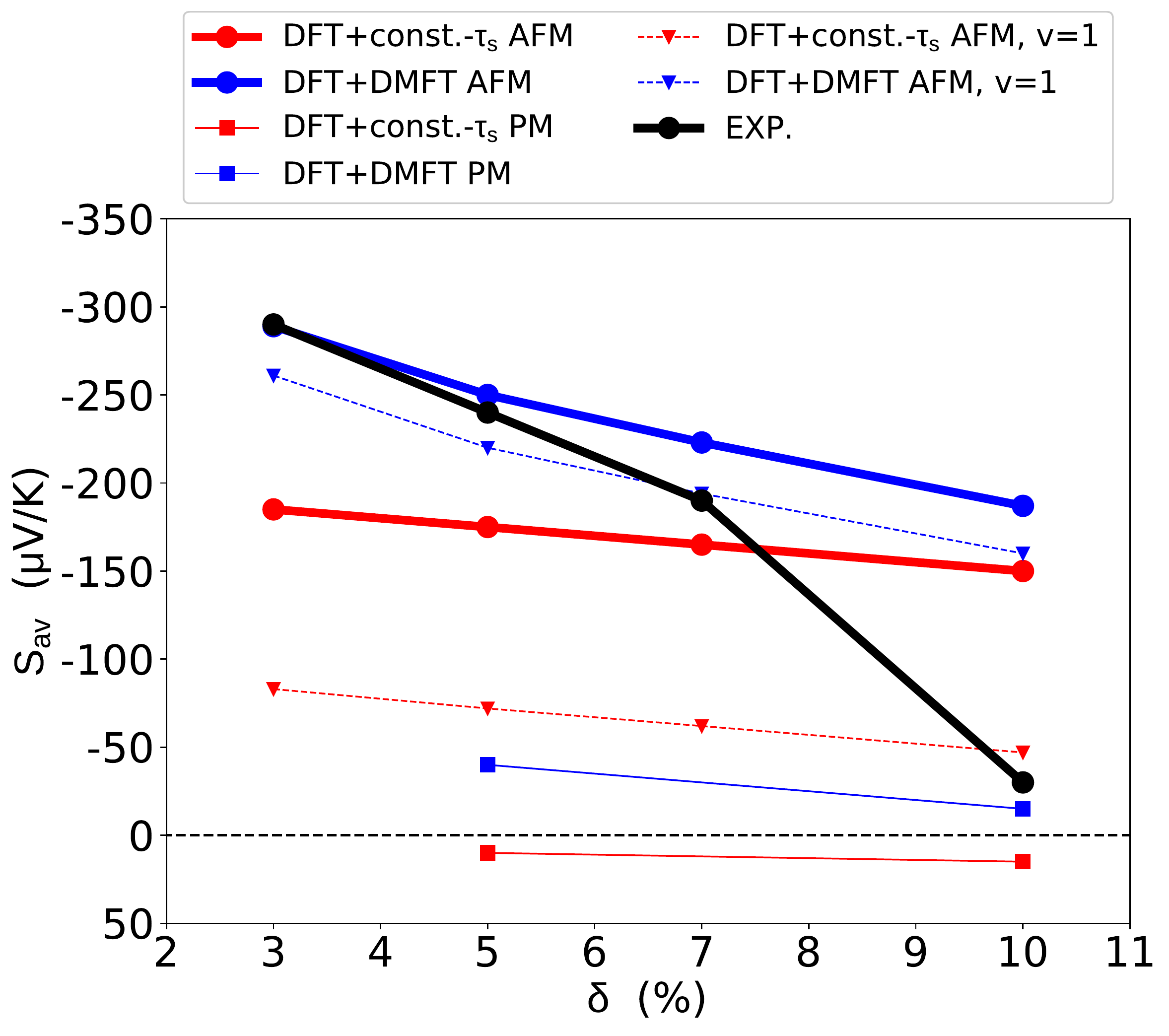}
	\caption{\label{fig:Sdope}Averaged Seebeck coefficient $S_{av}$
		as a function of electron doping level~$\delta$ calculated with
		spin-polarized (antiferromagnetic) DFT+const.-$\tau_s$ (red
		circles), DFT+DMFT in the antiferromagnetic phase (blue circles),
		non-magnetic DFT+const.-$\tau_s$ (red squares), DFT+DMFT in the
		paramagnetic phase (blue squares), and compared to
		experimental results from Ref.~\cite{Ryu} (black circles).
		The evaluation of $S_{av}$ with $v(\mathbf{k}) = \unit$ (Eq.~(\ref{eq:sv1}))
		in the antiferromagnetic phase is shown with blue and red triangles.
		The lines are a guide to the eye.} 
\end{figure}

Now, we turn to the doping dependence of the Seebeck
coefficient (Fig.~\ref{fig:Sdope}). The DFT+DMFT $S_{av}$ in the AFM phase is
\SI{-290}{\micro\volt\per\kelvin} at 3\% doping and increases up to
\SI{-190}{\micro\volt\per\kelvin} at 10\% doping (blue circles). At a
doping of 3\% the calculated value coincides with the experimental
data~\cite{Ryu} (black circles) and is still in a good agreement
at doping levels of 5 and 7\%. On the other hand, a paramagnetic (PM)
DFT+DMFT calculation (blue squares) of $S_{av}$ yields only
\SI{-40}{\micro\volt\per\kelvin} at 5\% doping. The
large discrepancy at 5\% between the PM and the AFM result suggest
that the magnetic ground state is an essential ingredient to describe the
thermoelectric properties of \ce{LaMnAsO} at the lower doping levels.  

The experimental data shows a strong change of $S$
to only $\SI{-30}{\micro\volt\per\kelvin}$ when doping is increased
from 7 to 10\%. A similar behavior has been observed for \ce{SmMnAsO_{1-$\delta$}}
samples~\cite{Sm01}. In this compound the Seebeck coefficient is
$S=\SI{-280}{\micro\volt\per\kelvin}$ for an oxygen-deficiency of
$\delta=0.17$ at room temperature, but upon a further increase
to $\delta=0.2$ it jumps to only $\SI{-40}{\micro\volt\per\kelvin}$.
In \ce{SmMnAsO_{1-$\delta$}} this change in $S$, which is also accompanied
by a strong increase of the conductivity, is connected to the transition to the PM state.
Although the N\'eel temperature in the case of \ce{F}-doped
\ce{LaMnAsO} has not been measured, the conductivity does change abruptly by
two orders of magnitude from 7 to 10\% doping~\cite{Ryu}. Furthermore, experiments
demonstrated that the AFM phase can be destructed under H doping of about 8-14\%~\cite{La03}.
At 10\% doping our calculation in the PM phase is in accordance with the experimental
value, which can be seen as a further hint that the suppression of $S_{av}$ from 7 to 10\%
doping is probably connected to the AFM-PM transition. However, in this work we do not
intend to investigate the phase transition from an AFM to a PM state within DFT+DMFT.
It is well know that DMFT is not too reliable in predicting the absolute value of
magnetic transition temperatures~\cite{BaLa_Zingl,Neel04, Neel06}, as well as the
transition as function of doping, which has been discussed for example in the
context of high-$T_C$ cuprate superconductors~\cite{Maier2005, Otsuki2014}.

The doping dependence of $S_{av}$, see Fig.~\ref{fig:Sdope}, shows that the
DFT+const.-$\tau_s$ approximation cannot provide an accurate description over
the full range of doping levels. For example, at 3\% doping $|S_{av}|$ is by
more than $\SI{100}{\micro\volt\per\kelvin}$ smaller than the
DFT+DMFT value and the experimental data. From the evaluation of
$S^{v=\unit}_{av}$ (red and blue triangles) we see that
the fundamental difference in the interpretation of $S_{av}$
within DFT+DMFT and DFT+const.-$\tau_s$, as discussed above at 5\% doping,
applies to the whole studied doping range. We finally point out that in contrast
to PM DFT+DMFT the non-magnetic DFT+const.-$\tau_s$ calculations result even in
a positive Seebeck coefficient of $\SI{10}{\micro\volt\per\kelvin}$ at 5\%
doping and $\SI{15}{\micro\volt\per\kelvin}$ at 10\% doping (red squares).

For thermoelectric applications not only $S$ is crucial, but more so the 
power-factor $S^2 \sigma$ (numerator of $ZT$). The calculated out-of-plane
conductivity $\sigma^{zz}$ of electron-doped \ce{LaMnAsO} is about a factor
50 lower in our DFT+DMFT calculations than the in-plane conductivity $\sigma^{xx}$.
This is a consequence of the quasi-two-dimensional nature of \ce{LaMnAsO}~\cite{BaLa_Zingl}.
The crystal symmetries demand $\sigma^{xx} = \sigma^{yy}$ and $S^{xx} = S^{yy}$, and thus
the averaged Seebeck coefficient is mainly determined by its in-plane value, $S_{av} \approx
S^{xx}$ (Eq.~(\ref{eq:sav})). However, we find that the quasi-two-dimensional nature
is not pronounced in the direction-dependent Seebeck coefficient itself. For
all studied doping levels in the AFM phase $|S^{zz}|$ is less than
\SI{40}{\micro\volt\per\kelvin} smaller than $|S^{xx}|$. Putting
everything together, the in-plane direction offers a slightly higher
Seebeck coefficient and exhibits a substantially higher conductivity than the
out-of-plane direction. Therefore, we predict that a possible
single-crystalline \ce{LaMnAsO_{1-$\delta$}F_$\delta$} sample should
show the highest power-factor ($S^2 \sigma$) if thermoelectricity is
harvested in the in-plane direction.  

\section{Conclusion}
\label{sec:Conc}

We studied the electronic influences on the Seebeck coefficient of
electron-doped \ce{LaMnAsO} within the framework of fully charge
self-consistent DFT+DMFT calculations. To model experimentally
synthesized \ce{LaMnAsO_{1-$\delta$}F_{$\delta$}} we used the
virtual crystal approximation at electron doping levels of $\delta=$
3, 5, 7 and 10\%. In DFT the doping pushes the bottom
of the flat $xz/yz$ bands below the Fermi energy. On the contrary,
the incorporation of electronic correlations within DMFT shows that doping
leads to incoherent weight at and below the chemical potential, whereas
the renormalization of the unoccupied states results in strongly-pronounced spectral
features located directly above it. Both DFT and DFT+DMFT
calculations predict negative Seebeck coefficients, however with
completely opposing underlying mechanisms. While the
DFT+const.-$\tau_s$ approximation points towards a picture where the
asymmetry in the velocities is pivotal, DFT+DMFT traces the Seebeck
coefficient almost exclusively back to the asymmetry of the correlated
spectral function. Therefore, our calculations demonstrate that the
interpretation of the Seebeck coefficient in materials with strong electronic
correlations and non-negligible incoherent spectral weight requires to go
beyond the constant scattering time approximation. Considering finite
life-time effects within the DFT+DMFT framework yields a higher Seebeck
coefficient in electron-doped \ce{LaMnAsO} than what would be anticipated
from DFT, and is also in much better agreement with experimental data.
The emergence of incoherent spectral weight under doping is unique to
correlated systems, and could potentially offer new routes in the
engineering of thermoelectric materials.

\begin{acknowledgments}
We thank R.~Triebl and J.~Mravlje for useful discussions.
The authors acknowledge financial support from the Austrian Science
Fund FWF (Y746, P26220, F04103) as well as NAWI Graz.  Calculations
have been performed on the Vienna Scientific Cluster (VSC).
The Flatiron Institute is a division of the Simons Foundation. 
\end{acknowledgments}

\appendix
\section{Virtual Crystal Approximation}
\label{sec:vca}

A simple way of incorporating the effect of doping within band-structure
methods is the virtual crystal approximation (VCA). Computationally,
the VCA is efficient, because calculations can be carried out at the
same cost as for the corresponding undoped structure. However, the VCA
neglects charge localization and assumes that there is a virtual atom on
all sites which interpolates between the original atom and the
dopant. This picture is only adequate for atoms with similar radii and
the same number of core electrons. Another possibility to take
doping effects into account is the construction of super-cells (SC),
where the doped atoms are directly replaced by the dopant in a larger
unit cell. Super-cells assume a long-range order of the dopants in the
crystal matrix. Using this approach within DFT+DMFT is certainly
feasible at high enough doping levels, but it would be demanding for
the lower doping levels used in this work due to the size of super-cells needed.

\begin{figure}[t!]
	\centering
	\includegraphics[width=1.0\linewidth]{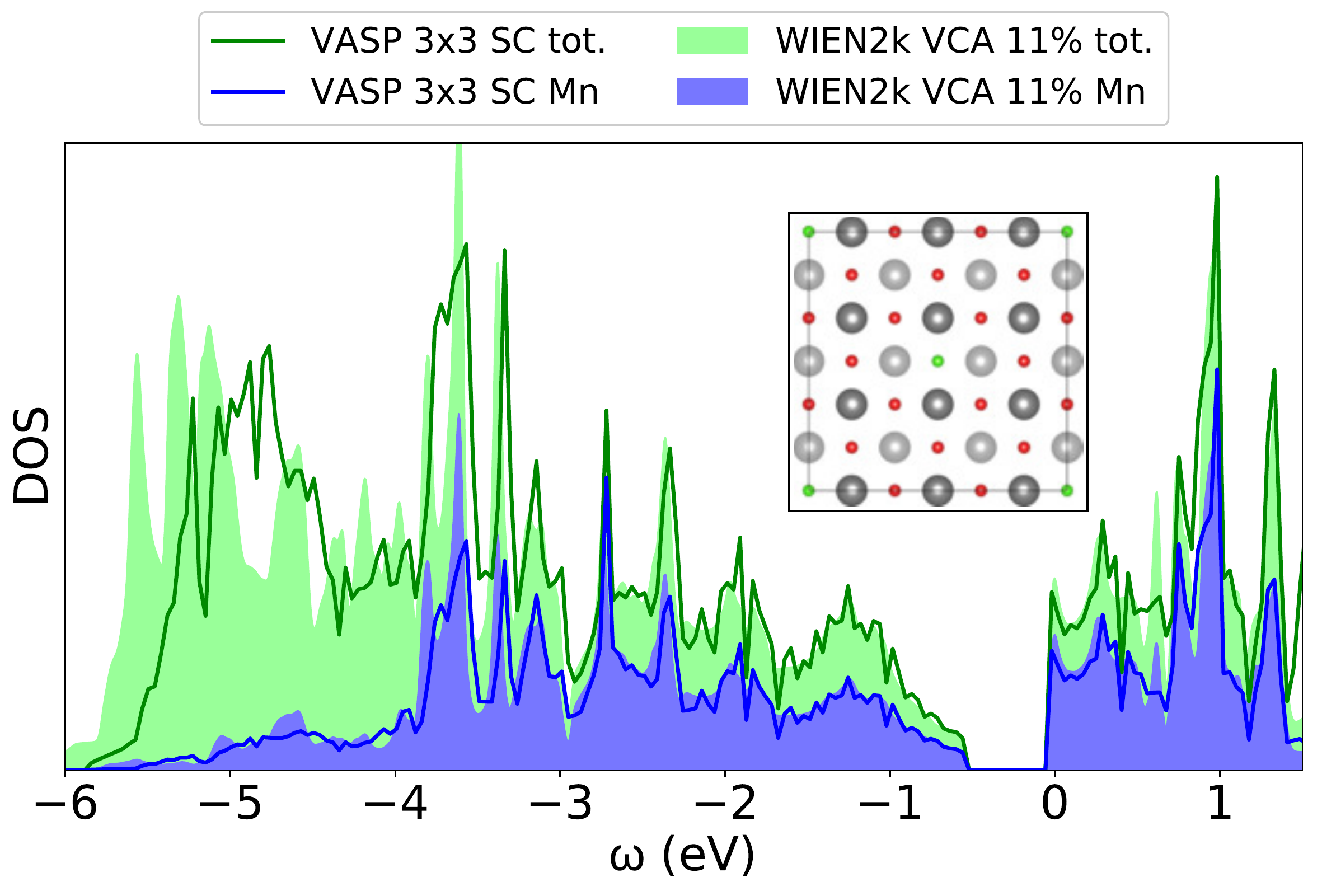}
	\caption{\label{fig:DFT_doping}Comparison of the WIEN2k VCA DOS
		in the antiferromagnetic state at 11\% electron doping (filled
		areas) with the VASP DOS for a $3\times 3 \times 1$
		super-cell (lines), i.e.\ a 11.11\% substitution of O with F. The
		total DOS is colored in green and the projected Mn-$3d$ DOS in
		blue. The inset (prepared with VESTA~\cite{VESTA}) shows a
		top view of the La-O layer, the positions of the F atoms (green)
		substituting the O atoms (red) are indicated. The La atoms sitting
		below the O plane are lighter colored than those above.} 
\end{figure}

To assess the applicability of the VCA to \ce{LaMnAsO_{1-$\delta$}F_{$\delta$}}
we compare our WIEN2k calculations to super-cell calculations carried out
with VASP 5.4.1~\cite{kresse_ab_1993,kresse_ab_1994, kresse_efficiency_1996, kresse_efficient_1996}
with the projector augmented wave (PAW) method~\cite{blochl_projector_1994, kresse_ultrasoft_1999}
and pseudopotentials v.54~\footnote{The pseudopotentials are: La 06Sep2000, Mn 06Sep2000, 
As 22Sep2009, O 08Apr2002, F 08Apr2002, Sr\_sv 07Sep2000.}. The plane wave energy cutoff
of is set to $\SI{400}{\electronvolt}$. Like in the WIEN2k calculations, the PBE density
functional and the same crystal structure parameters are used. The full BZ of the
super-cell is sampled with a $10\times10\times12$ $\Gamma$-centered
Monkhorst-Pack $\mathbf{k}$-grid~\cite{monkhorst_special_1976},
whereas for the WIEN2k calculation \num{10000} $\mathbf{k}$-points in
the BZ of the initial cell, which is 9 times smaller in real space,
are used. To be consistent with the experiment, we replaced the O
atoms by F atoms in a $3\times 3 \times 1$ super-cell, which corresponds
to a doping level of $11.\overline{1}\%$ (see inset of Fig.~\ref{fig:DFT_doping}).
In WIEN2k the VCA is employed by adjusting the atomic number of the O atoms
to $Z=8.11$. We note that the choice of WIEN2k for the VCA calculation and
VASP for super-cells is intrinsic to the differences in these two DFT codes.
In the full-electron code WIEN2k large super-cell calculations are demanding.
On the other hand, VASP is a pseudo-potential code, which makes it
cumbersome to use the VCA.

The agreement of the VCA with the super-cell calculations, Fig.~\ref{fig:DFT_doping},
is especially good in the energy region with no or only weak hybridization of the
La-O and Mn-As layers, which is roughly between \SI{-3.5}{\electronvolt} and
\SI{1.5}{\electronvolt}. Note that the Mn-$3d$ projected DOS are in even better
agreement than the total DOS. The former is the more important quantity as only
the Mn-$3d$ orbitals are treated within DMFT. Of course, in energy regions
exhibiting dopant states one cannot expect a good agreement between the VCA
and the super-cell calculation. This is visible from \SI{-6.0}{\electronvolt}
to about \SI{-3.5}{\electronvolt}, where the DOS is mainly determined by
O states, i.e.\ the properties of the La-O layer.  

Super-cell calculations with a different arrangement of the dopants in
the unit cell did not substantially change the DOS. We also compared
the VCA to a super-cell calculation for the non-spin-polarized state
and found an agreement on the same level as for the spin-polarized
calculations (not shown). Furthermore, calculations for the undoped compound
gave perfect agreement between WIEN2k and VASP. As we are mainly interested
in spectral properties in the vicinity of the Fermi energy, which are to a great
extent determined by the Mn-As layer, the comparison presented in
Fig.~\ref{fig:DFT_doping} underlines that VCA is an eligible approximation
for the doping levels considered in this study.
~~\\
~~\\
\bibliography{literatur.bib}
\end{document}